\documentclass[11pt,a4paper]{amsart}
\usepackage[dvips]{graphicx}
\usepackage{amsmath}
\newcommand\citen[1]{\cite{#1}}
\newcommand{\ket}[1]{{|#1\rangle}}
\newcommand{\bra}[1]{{\langle #1|}}
\newcommand{\vect}[1]{{\mathbf{#1}}}
\newcommand{\cur}[1]{{\mathcal{#1}}}
\begin{document}
\title{Classical Roots of the Unruh and Hawking effects}
\author[M.~Pauri and M.~Vallisneri]{M.~Pauri$^1$ \and M.~Vallisneri$^2$}
\begin{abstract}
Although the Unruh and Hawking phenomena are commonly linked to field quantization in ``accelerated'' coordinates or in curved spacetimes, we argue that they are deeply rooted at the classical level. We maintain in particular that these effects should be best understood by considering how the special-relativistic notion of ``particle'' gets blurred when employed in theories including accelerated observers or in general-relativistic theories, and that this blurring is an instantiation of a more general behavior arising when the principle of equivalence is used to generalize classical or quantum special-relativistic theories to curved spacetimes or accelerated observers. A classical analogue of the Unruh effect, stemming from the non-invariance of the notion of ``electromagnetic radiation'' as seen by inertial and accelerated observers, is illustrated by means of four {\it gedanken-experimente}. The issue of energy balance in the various cases is also briefly discussed.
\end{abstract}
\maketitle
\footnotetext[1]{Dipartimento di Fisica, Universit\`a di Parma, 43100 Parma, Italy; INFN, Sezione di Milano, Gruppo Collegato di Parma, Italy. E-mail address: pauri@parma.infn.it}
\footnotetext[2]{Theoretical Astrophysics 130-33, Caltech, Pasadena CA 91125; Dipartimento di Fisica, Universit\`a di Parma, 43100 Parma, Italy; INFN, Sezione di Milano, Gruppo Collegato di Parma, Italy. E-mail address: vallis@caltech.edu}
\setcounter{footnote}{2}
\section{Introduction}
\noindent The Unruh and Hawking effects sit deservedly among the most widely discussed and popularized subjects in the physics of the last two decades. A strong part of their ``folklore'' is the conviction that these effects have an eminently \emph{quantum mechanical} character. For instance, one often hears that black holes would indeed be \emph{black} by classical physics, were it not for quantum mechanics coming to the rescue of black hole thermodynamics, providing a thermal emission of particles from the black hole's horizon (\emph{Hawking effect} \cite{Hawking75}). And again, the fact that the Minkowski vacuum should contain particles to be seen by an accelerated detector (\emph{Unruh effect} \cite{Unruh76}) is perceived as a modern quantum marvel \emph{on par}, say, with quantum tunneling and EPR effects\footnote{As insightfully discussed by Sciama \cite{Sciama79}, these phenomena bring together in an intriguing way Einstein's independent legacies of fluctuation theory and relativity.}. In this paper we claim instead that both the Unruh and the Hawking effect have a clear classical counterpart, and that they can be understood as typical examples of the \emph{perspectival semantics} arising within the difficult migration from special-relativistic to curved-spacetime physics or simply to accelerated observers \cite{Vallisneri97}.

The assertion of the Poincar\'e group as the global symmetry group of space-time has been seminal to the great theoretical synthesis of the first half of this century, begun with the full acknowledgment of Maxwell's electro-magnetism as a special-relativistic theory, and beautifully climaxed with quantum field theory. Thus, the concepts and interpretive paradigms of these theories refer naturally to the privileged class of inertial observers. Now, the equivalence principle of general relativity does warrant Lorentz group as a symmetry group, but only locally: this locality becomes crucial when one tries to generalize to curved space-time geometries the concepts and paradigms inherited from special-relativistic theories, when these are based on the global symmetries of Minkowski space-time.

In section \ref{sec:classicuh} of this article we shall argue that the gist of the Unruh and Hawking effects can be understood within this frame of reasoning, well beyond their quantum character. Essentially, we shall be discussing how the special-relativistic notion of a quantum ``particle'' becomes \emph{slippery} when one tries to extend it to curved spacetimes or to non-inertial observers.

In section \ref{sec:equivalence} we will show that the \emph{same} ambiguity befalls the entirely classical concept of electro-magnetic ``radiation'', by examining the especially instructive ``paradox'' of a charge falling in a constant homogeneous gravitational field: by emitting radiation, such a charge might be distinguishable from a similarly falling uncharged body, pointing to a violation of the equivalence principle of general relativity. We will deliberately introduce the issue in a blurred way echoing its initial appreciation in the literature, as a borderline case between special and general relativity; this presentation makes the contradiction most apparent. By fully placing the question within the theoretical framework of general relativity, the ``paradox'' is seen to fade. The solution lies in the fact that the notion of electro-magnetic radiation \emph{is not invariant} with respect to transformations between inertial and accelerated reference frames, so that radiation can be ``produced'' or ``transformed away'' by suitably changing the state of motion of the observer.

We regard this ``illusion'' as a veritable forerunner of the Unruh and Hawking effects and we submit that these effects are, in R.\ Peierls' definition \cite{Peierls79}, ``intellectual surprises'' that could have been foreseen much earlier, were it not for the difficult epistemic \emph{upgrade} required to switch from the special to the general-relativistic worldview.
\section{Classical nature of the Unruh and Hawking effects}
\label{sec:classicuh}

\subsection{The Unruh and Hawking effects: synopsis}

As many authors have underlined, the Unruh and Hawking effects are most transparently explained as being about the \emph{different definitions} of ``quantum particles'' from alternative points of view.

An essential ingredient in the standard quantization of free fields theories is the normal mode decomposition of the field operators and the distinction between ``positive frequency'' and ``negative frequency'' modes, fixing the identity of ``particles'' and ``antiparticles'' and, most importantly, that of the ``vacuum state''.
However, there are infinitely many ways to accomplish this decomposition\footnote{See for instance ref.~\citen{Wald94}.}, roughly corresponding to all the possible choices of a complete set of solutions for the classical wave equation. While building a correspondence between such differently assembled theories\footnote{Even if different choices of the modes may in general lead to \emph{unitarily inequivalent} theories \cite{Wald94}, it is always possible to establish an arbitrarily ``accurate'' correspondence between the states of any two such constructions, using the so-called \emph{algebraic approach} \cite{Haag64}.}, cases are found when the vacuum state of a theory is mapped to a ``particle''-containing state in another theory.

In special-relativistic theories, there is a strong criterion to select one particular quantization: one traditionally picks classical solutions of definite frequency with respect to Minkowski coordinate time, thereby insuring a covariant notion of ``particle'' that is adequate for all inertial observers. If we extend our scope, however, we soon discover that there can be more than one logical choice of modes.

A first example is the Unruh effect \cite{Unruh76}, which takes into account an observer travelling through Minkowski space-time along a uniformly accelerated world-line. To him it will come natural to employ modes of definite frequency with respect to his \emph{proper time}: in this way the vacuum state of the usual theory based on Minkowski coordinate time is found to correspond to a thermal bath of ``particles'' in the theory built starting from ``accelerated modes''.

This result is considered ``robust'' since it can be derived by an altogether different approach \cite{Unruh76}: by standard approximation theory, a quantum detector moving along a pre-determined accelerated world-line is found to thermalize upon interaction with the ``Minkowski'' vacuum state.
It is a well-known result that its energy absorption rate is essentially determined by the Fourier transform of the \emph{autocorrelation function} of the field, along the detector's world-line and with respect to the detector's proper time:
\begin{equation}
R(\omega)=\int^{+\infty}_{-\infty} d\tau\, e^{-i \omega \tau}
\bra{0}{\hat{\phi}}(x^{\mu}(0)){\hat{\phi}}(x^{\mu}(\tau))\ket{0}.
\label{eq:rate}
\end{equation}
Since the Wiener-Khinchin theorem states that the Fourier-transformed autocorrelation of a signal is actually its power spectrum, the detector response is seen to be correlated to the energy content of the field in a manner that is dependent on the specific form of the energy-momentum tensor of the field, and of the interaction Hamiltonian between the field and the detector.
This substantiates the presence of a thermal signal, which is readily interpreted in the light of the ``normal mode'' derivation of the Unruh effect as belonging to a thermal bath of ``particles''. Any absorbed energy must ultimately come from the classical agency supposed to keep the detector on its worldline.

Moving to curved space-times, consider the Hawking effect \cite{Hawking75}: here a quantum field is taken to live on the background geometry of a spherically symmetric distribution of matter collapsing to a black hole. Now, the symmetries of this space-time hint to two natural definitions of quantum ``particle'': one is appropriate to observers inhabiting the early stages of collapse, when space-time is still approximately Minkowskian; the other suits late observers witnessing the stationary black hole phase. It turns out that the vacuum state, as defined by early observers, appears to late ones to contain a thermal bath of particles incoming from the direction of the black hole's event horizon. Again, this conclusion is fortified by consideration of quantum detectors traveling through Schwarzschild space-time \cite{Vallisneri97}.

\subsection{Slippery notions and interpretive illusions}

A physical theory consists loosely of three interpenetrating bodies of knowledge: an axiomatic structure identifying principles, laws and the consequences that can be inferred from them; an \emph{operative} interface to experimentation, together with a set of defining or \emph{encyclopaedic} experimental results; and a \emph{semantic} framework of interpretations and paradigms necessary to conceive the physical world and to think about physical facts.
As an example, for the standard ``one-particle'' quantum theory the axiomatic construction would essentially be that of Dirac's \emph{Quantum Mechanics} \cite{Dirac58}, while the interpretive framework might be identified with the ``Copenhagen interpretation''.

Each of these three segments has its ways of restructuring and evolving; predictably, the semantics of a theory are the fluidest, often depending on unspoken perceptions and understandings, and rarely residing organically in written documents.
Some of them they crystallize into what we call ``notions'': some of them are doomed to extinction (think of the ether); others pass unscathed or even augmented from a successful theory to the next one (think of mass and energy); and again others are subject to curious blurrings and cross-breedings (think of ``particles'' and ``waves'' after the quantum revolution).
We regard this ``memetics'' \cite{Dawkins76} of notions as one of the most charming and enjoyable versants of the history of theoretical physics.

We will say that we are in presence of \emph{perspectival semantics} when within a theory the same physical \emph{information} is assigned distinct \emph{semantic contents} according to different but equally legitimate readings.
This happens for quantum field theory when it is tentatively extended to curved space-times, or to accelerated observers. Even if Einstein provided the principle of equivalence to ferry special-relativistic physics across to general relativity, the old semantics cannot always cope with the upgrade: some notions get \emph{slippery}, or become afflicted with paradoxes.

The Unruh and Hawking effects are instantiations of perspectival semantics wherein the ambivalent signal is the value of the field, and its perspectival interpretation points to the failure of the notion of ``particle''. 
Let us then dissect this very notion. We feel entitled to speak of quantum ``particles'' when we remark a certain \emph{periodic} structure in the temporal and spatial dependence of the field signal. This attribution of meaning is not new to quantum field theory, but can be traced etymologically to certain basic tenets of \emph{parent theories}:
\begin{enumerate}
\item in \emph{one-particle quantum mechanics}, solutions to the wave equation are seen as describing, once again, a ``particle'';
\item in \emph{classical non-relativistic mechanics}, position and momentum are \emph{fundamental observables}, fully defining the location of the representative point in phase space; 
\item the passage to \emph{relativistic classical mechanics} somehow weakens the fundamental status of the position observable, due to covariance problems; on the other hand, the energy-momentum four-vector gains clout as the ``essential attribute'' of a relativistic particle;
\item again \emph{in relativistic classical mechanics}, energy and momentum are the generators of infinitesimal translations in time and space; Fourier modes are then identified as waves (and, consequently, ``particles'') of definite energy-momentum.
\end{enumerate}
By the covariance properties of the energy-momentum four-vector, all \emph{inertial} observers in \emph{Minkowski} space-time will perform \emph{compatible} frequency analyses of the same signal, leading to coherent identifications of ``particles''.
In the Unruh and Hawking effects, a ``particulate'' content is ascribed to states, otherwise seen as ``empty'', by means of a frequency assessment  which however takes place outside of the \emph{compatibility domain} of the ``particle'' notion, i.~e.\ special-relativity. Indeed, the Unruh and Hawking effects come about when we try to enlarge quantum field theory to accommodate accelerated observers in Minkowski space-time or generic observers in curved space-times.

\subsection{The Unruh and Hawking effects in classical field theory}

We come now to our claim. The normal mode decomposition of the field arguably belongs to the \emph{classical} domain: for instance, in classical field theory one can write a real scalar field as a sum of a complete set of positive frequency orthonormal modes,
\begin{equation}
\phi(x^{\mu}) = \sum_i a_i \psi_i(x^{\mu}) + a_i^{*} \psi^{*}_i(x^{\mu}),
\label{eq:decompa}
\end{equation}
and interpret the coefficients $a_i$'s and $a_i^{*}$'s as articulating the presence of single wave-modes in the overall configuration of the field. Quantum field theory is obtained by ``promoting'' these coefficients to Fock algebra operators. The ``particle content'' of a quantum state is then ``read'' by means of the number operators $N_i \equiv a_i^{\dagger} a_i$.

If two competing decompositions are set up for the field, as both in the Unruh and Hawking effects, the transformation between the coefficients (operators) in the two schemes \emph{will not depend on the procedure (classical or quantum) employed to ``read'' the field signal}, but only on the way in which one set of modes can be written in terms of the other: namely, on their reciprocal scalar products\footnote{Note that this is true already at the classical level. The usual way to define scalar products in free quantum field theories is to adapt the \emph{symplectic structure} of the space of solutions of the classical wave equation (see e.~g.\ ref.~\citen{Wald94}). This procedure insures the conservation of scalar products through evolution.
At the classical level, these scalar products can be used to set up for any solution a spectral decomposition that will be conserved in time: i.~e., any solution can be seen as a superposition of wave-modes.

Once a second decomposition ``competing'' with \eqref{eq:decompa} is established as
\begin{equation*}
\phi(x^{\mu}) = \sum_i c_i \xi_i(x^{\mu}) + c_i^{*} \xi^{*}_i(x^{\mu}),
\label{eq:decompb}
\end{equation*}
we obtain for the new coefficients and for the creation and destruction operators
\begin{gather*}
c_j = \sum_i \alpha_{ij} a_i + \beta_{ij}^{*} a_i^{*} \\
\alpha_{ij} = (\xi_j,\psi_i), \hspace{0.5cm} \beta_{ij} = (\xi_j^{*},\psi_i),
\end{gather*}
where the scalar product is the one defining the $\xi_i$'s as a complete orthonormal set.}. As we have already remarked, the gist of the Unruh and Hawking effects resides in this transformation, which we now acknowledge as originally classical. What is more, equation \eqref{eq:rate} closely parallels the expression found by Planck \cite{Planck00} for the rate at which a classical charged harmonic oscillator absorbs energy from a statistical radiation field:
\begin{equation}
R_{\mathrm{cl}}(\omega)=\int^{+\infty}_{-\infty} d\tau\, e^{-i \omega \tau}
\langle \phi(x^{\mu}(0))\phi(x^{\mu}(\tau)) \rangle,
\end{equation}
where $\omega$ is the oscillator's natural frequency and $\langle \ldots \rangle$ denotes an \emph{ensemble} average.

Does then classical field theory exhibit the Unruh effect? It does not if we consider its ``fundamental'' configuration to be given by an everywhere null field, which is truly a universally invariant configuration! No matter how a null signal is read, it will always remain null. In quantum field theory, instead, the non-vanishing fluctuations of the vacuum state always provide a bias signal that makes the Unruh and Hawking \emph{perspectival effects} possible.

There are two ways to introduce such a ``fundamental signal'' in classical field theory: the first is simply to bring in classical sources, and to examine the wave-mode content of the resulting inhomogenous solutions of the wave equation. Higuchi and Matsas \cite{Higuchi93} define a ``classical particle number'' as energy density per unit frequency divided by frequency, and proceed to show that the relation between these numbers, as reckoned in inertial and ``accelerated'' coordinates, is consistent with the existence of an Unruh-type thermal bath. A related approach is due to Srinivasan \emph{et al.} \cite{Srinivasan97a,Srinivasan97b}

The second route is to postulate that the ``fundamental'' configuration of the classical field consists of an incoherent superposition of plane waves, endowing the vacuum with a ``zero-point'' energy of $\hbar \omega / 2$ per mode; the wave phases are assumed to be uniformly and independently distributed random variables.
This is called \emph{stochastic classical field theory}; in its specialization to electro-magnetism it was introduced in 1963 by Marshall \cite{Marshall63,Marshall65}; the constant $\hbar$ is imported to the classical framework by requiring the mean-square displacement of a charged harmonic oscillator to be the same as in quantum theory. The ``fundamental signal'' of stochastic classical field theory is able to account for the Unruh effect \cite{Boyer84}.

It is arguable that all the above derivations leading to the appearance of the Unruh effect in classical field theory could be reproduced for fields inhabiting the background space-time of a gravitational collapse, engendering a classical Hawking effect. This conviction stems from the circumstance that both effects are derived, as we discussed above, by the essentially classical operation of tracing ``competing'' sets of modes and calculating their reciprocal scalar products. We defer to a future paper the explicit calculations.

It is interesting to ponder whether these classical homologues could have been noticed during the development of classical electro-magnetism, so that the Unruh and Hawking effects would have subsequently been derived as their ``quantum version''. The answer is probably negative. These classical results have a distinct retrospective flavor, due in part to the weaker semantic content of the notion of ``wave-mode'' compared to that of ``particle'', and in part to the absence of a fluctuating vacuum to highlight the phenomenon in the classical domain.

In spite of this, we believe that the Unruh and Hawking perspectival effects could have been predicted earlier and by a different route, drawing from the analogy with the slippery notion of ``radiation'' in classical electro-magnetism, when this theory is tentatively extended to general relativity. \emph{Slippery radiation} is the subject of the next section.
\section{The equivalence principle paradox}
\label{sec:equivalence}
\begin{figure}
\begin{center}
\includegraphics[width=5in]{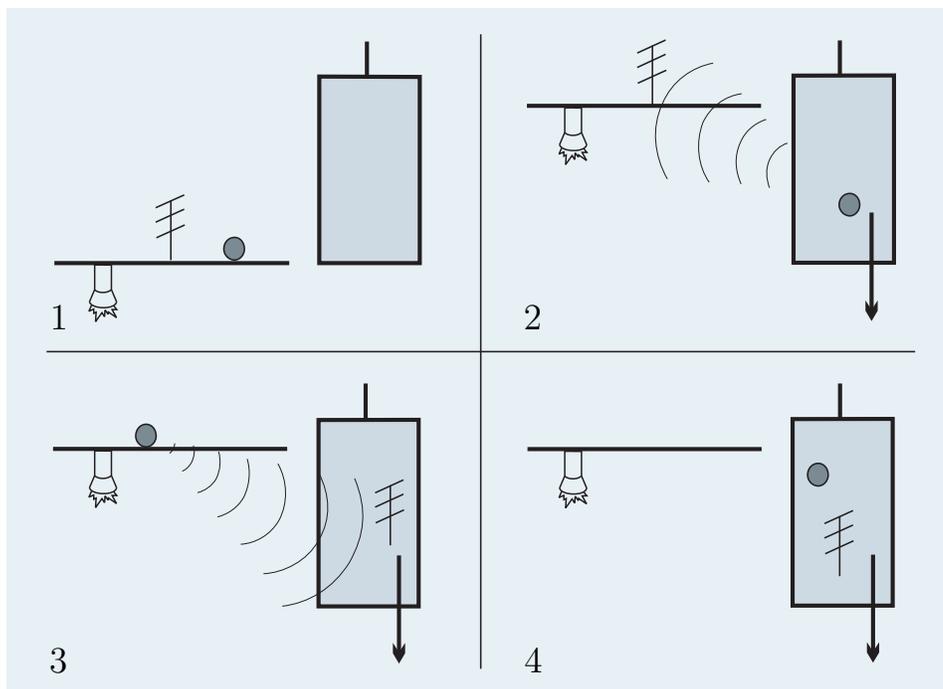}
\caption{Four \emph{Gedankenexperimente}: To the right of the laboratory frame, \emph{supported} by a compensating agency (rockets), is our imaginary \emph{Einstein's elevator}, falling freely (except for experiment 1) in the earth's gravitational field, which we assume to be homogeneous.
\label{fig:tests}}
\end{center}
\end{figure}

\noindent One of the challenges posed by the advent of general relativity to the established comprehension of the physical world was the apparent conflict between the principle of equivalence and the well established fact that accelerated charges radiate. This can be spelled out by the following \emph{Gedankenexperiment}: let us move to a laboratory setting here on Earth and do tests with a system consisting of a point-like electric charge and a detector of electro-magnetic radiation. We will check whether the detector registers any radiation when the system is set up as follows (see fig.\ \ref{fig:tests}):
\begin{enumerate}
	\item support both the charge and the detector in the Earth's gravitational field;
	\item support the detector and let the charge fall freely;
	\item let the detector fall and support the charge;
	\item let both the detector and the charge fall freely.
\end{enumerate}
If we are willing to concede that our laboratory is small enough compared to the Earth, we may work in the idealization that the detector and the charge are immersed in a \emph{constant homogeneous gravitational field}: falling objects will then move along uniformly accelerated trajectories (possibly relativistic) in the ``vertical'' direction.

Let us first consider experiments 1 and 2. Our pre-relativistic intuition suggests that the still charge will emit no radiation, while the falling one will; what is more, since the falling charge will lose energy to electro-magnetic radiation, it will fall more slowly than a similar uncharged body. Yet the \emph{equivalence principle} of general relativity, at least in the case of homogeneous (\emph{apparent}) fields, requires a charged \emph{test} particles to follow the same geodesics as an uncharged one\footnote{Of course, since our charge is still a \emph{test}-particle, one works under the assumption that neither the charge's mass nor that of the electromagnetic field it generates are significant to the shaping of space-time geometry.}!

By 1960, however, the very existence of radiation from a uniformly accelerated charge was still disputed. In V.\ Ginzburg's words \cite{Ginzburg70}, this is one of the ``perpetual problems'' of classical electrodynamics, and its discussion continued for decades in the scientific literature. M.\ Born's original solution for the field of a uniformly accelerated charge \cite{Born09} was divergingly interpreted as implying the emission of radiation \cite{Schott12,Schott15,Milner21,Drukey49,Bradbury62,Leibovitz63,Grandy70} or its absence \cite{vonLaue19,Pauli21,Rosen62}. Most notably, in his 1921 \emph{Enzyklop\"adie der Matematischen Wissenschaften} article, Pauli ruled that uniformly accelerated charges do not radiate.

In section \ref{subsec:uniform} we briefly summarize the debate and see that uniformly accelerated charges \emph{do} radiate according to the standard Larmor's formula,
\begin{equation}
\cur{R} = \frac{2}{3} \frac{e^2 a^2}{c^3}.
\label{eq:larmor}
\end{equation}
Once the presence of radiation is established, we are left with our ``paradox'', attracting by itself an extensive literature \cite{Bondi55,Fulton60,Rohrlich61,Rohrlich63,Mould64, Kovetz69,Ginzburg70,Boulware80,Piazzese85}. For instance, it has been argued \cite{Bondi55,Fulton60} that measurements of radiation may be made only in the limit of large distances from the falling charge, and that by various considerations it is doubtful that such an extended homogeneous gravitational field could be confidently postulated.

Yet a riper resolution may be found by framing the issue correctly within our modern understanding of general relativity \cite{Rohrlich63,Kovetz69,Ginzburg70}. The (strong) principle of equivalence \cite{Weinberg72,Misner73,Ciufolini95} can be formulated as stating that \emph{the special-relativistic equations of physics are valid unmodified in (local) inertial reference frames}. Coming to our experiments, in the hypothesis of a homogeneous gravitational field, Maxwell's special-relativistic equations are valid \emph{globally} throughout space-time, but only when they are written in the \emph{freely falling reference frame}.

Hence, we realize that Larmor's formula \eqref{eq:larmor}, together with the principle of equivalence, should not be used to predict the outcome of the ``supported'' experiments 1 and 2, but rather that of 3 and 4. The freely falling system consisting of detector and charge will behave exactly as would a similar system at rest, were gravitation absent, so that experiment 4 will report no radiation; whereas in experiment 3 the supported charge will be arguably accelerated relative to the freely falling detector, emitting radiation as given by \eqref{eq:larmor}. This is not a violation, but a \emph{consequence} of the equivalence principle!

Yet, if these results held true also for the supported \emph{detector}, we would be led to accept that, contrary to our earlier intuition, charges that are accelerated relative to the laboratory reference frame (experiment 2) do not radiate, while still charges (experiment 1) do. This latter conclusion would be especially embarrassing, since it is not clear how a continuous transfer of energy could be obtained within a \emph{stationary} physical system such as that of experiment 1. 

The point here is that the extension of outcomes obtained in the inertial frame to the ``supported'' experiments is unwarranted. We cannot infer the readings of the supported detector from those of the inertial one, but we must explicitly derive them within a suitable \emph{extension} of the special-relativistic theory of electro-magnetism. There are several ways to do so: by modeling a simple radiation detector and examining its response to electro-magnetic fields while undergoing acceleration \cite{Mould64}; by using a weak field approximation to general relativity \cite{Kovetz69}; or by evaluating the flux of the Poynting vector through spherical surfaces at rest in the ``supported'' frame\cite{Piazzese85}.

Following Rohrlich \cite{Rohrlich63}, in \ref{subsec:frame} we shall instead seek a set of coordinates for the Minkowski metric that may be taken to ``simulate physics'' as seen by observers supported in a constant homogeneous gravitational field. In this scheme, the outcome of experiments 1 and 2 is predicted by transforming to ``supported coordinates'' the field tensors found in the inertial system for supported and freely falling charges. In \ref{subsec:tests} we will see that in the supported frame the field of the still charge is very nearly of Coulomb form (experiment 1); while the falling charge is found to emit radiation (experiment 2).

By this procedure and in accord with refs.~\citen{Mould64,Kovetz69}, the results that we intuitively expected from the ``supported'' experiments are seen to be correct to a very good approximation for ``reasonable'' gravitational accelerations\footnote{See equation \eqref{eq:modifiedcoulomb}.}. Also, it is found that \emph{the very notion of radiation is not invariant with respect to transformations from inertial to accelerated reference frames}: as remarked in ref.~\citen{Sciama79}, much like ``virtual particles'' for the Unruh effect, \emph{an accelerated} (in this case, supported) \emph{observer will detect radiation where a freely falling observer only sees a pure Coulomb field}\footnote{Experiment 2 is the classical analog of the quantum Unruh effect, while experiment 4 would correspond to the quantum statement that an inertial detector sees no ``particles'' in the Minkowski vacuum. Levin, Peleg and Peres \cite{Levin92} discuss what might be considered a quantum analog of experiments 1 and 3. They introduce a quantum field living within a uniformly accelerated cavity and study its interactions with comoving or inertial detectors.}. Also, as in the Unruh effect, the energy absorbed by the accelerated observer must ultimately come from the agency enforcing acceleration, and not from the putative source of radiation. This is discussed in subsection \ref{subsec:energy}.
\subsection{The radiation of uniformly accelerated charges}
\label{subsec:uniform}
\emph{Uniformly accelerated motion} is defined in special relativity by requiring that the world-line $x^{\mu}(\tau)$ have a 4-acceleration $a^{\mu}=d^2{x^{\mu}} /{d \tau^2}$ of constant norm $(a^{\mu}a_{\mu})^{1/2}=g$, or equivalently that the 3-acceleration $\vect{a}(\tau)$ be a constant vector in the instantaneous rest frame of the world-line. If we restrict our attention to motions taking place on a 2-dimensional space-time plane\footnote{Relaxing this hypothesis yields the larger class of \emph{Synge helixes} \cite{Synge67}.}, we get (up to Poincar\'e transformations) the world-line \cite{Pauli21,Rohrlich65,Misner73}
\begin{equation}
\left\{
\begin{aligned}
t &= g^{-1}\sinh g\tau, \\
x &= 0, \\
y &= 0, \\
z &= g^{-1}\cosh g\tau.
\end{aligned}
\right.
\label{eq:hyperbolic}
\end{equation}
Since these equations describe a hyperbola in the $zt$ plane, this kind of motion is also known as ``hyperbolic'', in contrast with the ``parabolic'' free fall of Galileian mechanics. The reference frame that we have employed to write \eqref{eq:hyperbolic} is actually the \emph{instantaneous rest frame} of the moving point at proper time $\tau=0$. The trajectory turns out to be invariant with respect to Lorentz boosts on the $z$ axis, which shift it along itself; the boosts actually amount to simple translations in proper time transforming between instantaneous rest frames at different proper times.

Is \eqref{eq:hyperbolic} also a correct worldline for a charged particle coupled to the electro-magnetic field? It appears to be so, since substituting it in the standard Dirac-Lorentz equation \cite{Dirac38,Jackson62,Rohrlich65},
\begin{equation}
ma^{\mu} - F_{\mathrm{ext}}^{\mu} = 
\frac{2}{3} \frac{e^2}{c^3}  \left( \dot{a}^{\mu} - \frac{a^{\alpha}
a_{\alpha}}{c^2}
u^{\mu} \right ),
\label{eq:lorentz}
\end{equation}
yields a vanishing radiative damping term, so that \eqref{eq:hyperbolic} solves \eqref{eq:lorentz} for a suitable external field $F_{\mathrm{ext}}^{\mu}$. This circumstance has been the root of many misgivings: because the charge apparently loses no mechanical energy to radiation, it seems natural to conclude that there is no radiation at all. We will come to this in a moment.

The explicit form of the fields associated with hyperbolic motion was first derived by Born \cite{Born09}, and may be expressed \cite{Fulton60} in the usual cylindrical coordinates $(t,\rho,\phi,z)$ as
\begin{equation}
\left\{
\begin{aligned}
E_{\rho} &= 8 e g^{-2} \rho z / \xi^3, \\
E_{\phi} &= 0, \\
E_z &= -4 e g^{-2} (g^{-2} + \rho^2 + t^2 - z^2) / \xi^3, \\
H_{\rho} &= 0, \\
H_{\phi} &= 8 e g^{-2} \rho t / \xi^3, \\
H_z &= 0,
\end{aligned}
\right.
\label{eq:fieldcomp}
\end{equation}
where $\xi = [(g^{-2} - \rho^2 + t^2 - z^2)^2 + 4 g^{-2} \rho^2]^{1/2}$. In the hypothesis of \emph{retarded potentials}, the fields must be restricted to the \emph{causal future} $z + t > 0$ of the charge. This condition was not enforced in Born's original solution and was introduced by Schott \cite{Schott12,Schott15}. Bondi and Gold \cite{Bondi55} further ``patched'' the solution with Dirac-delta fields on $z + t = 0$, where otherwise it would not satisfy Maxwell's equations.

The magnetic field (and thus Poynting's vector) vanishes throughout space at time $t=0$. By symmetry, it must also vanish in every instantaneous rest frame, at all events that are simultaneous with the charge in that frame. Pauli \cite{Pauli21} concludes that ``there is no formation of a wave zone nor any corresponding radiation''.

Now, the notion of ``electro-magnetic radiation'' is usually associated with two connected physical facts:
\begin{enumerate}
	\item the fields originating at an event along the world-line of the charge, and propagating outwards on the light cone, consist both of a Coulomb term decreasing as $1/R^2$ (where $R$ is the radius of the light cone in any given Lorentz frame) \emph{and} of a $1/R$ \emph{radiation term} that eventually comes to dominate the field in the so-called \emph{wave zone};
	\item the \emph{radiation term} arises because at successive instants, the accelerating charge is not in the right position to support its previous Coulomb field. Thus a portion of field is effectively \emph{splintered away}: it takes on an independent existence and travels outwards from the charge, at the speed of light, carrying its own endowment of energy-momentum.
\end{enumerate}
We may then adopt, following Fulton and Rohrlich \cite{Fulton60}, a \emph{local, Lorentz-invariant criterion} to decide if a charge is \emph{instantaneously} radiating at event $x^{\mu}(\tau)$ along its world-line by evaluating the flux of the energy-momentum tensor $T^{\mu \nu}$ through light spheres centered at $x^{\mu}(\tau)$. In the limit of radii increasing to infinity\footnote{In order for the limit to be finite, the field must have a $1/R$ asymptotic behavior, since $T^{\mu \nu}$ is quadratic in $F^{\mu \nu}$.} we get a unique 4-vector, given in terms of the kinematic parameters of the charge's trajectory as
\begin{equation}
\frac{d P^{\mu}}{d \tau} = \frac{2}{3} \frac{e^2}{c^3} (a^{\alpha}
a_{\alpha})u^{\mu}.
\label{eq:criterion}
\end{equation}
According to this criterion, the uniformly accelerated charge is seen to radiate with Larmor's power $\cur{R} = 2/3 \, e^2 g^2 / c^2$. As for Pauli's objection, the vanishing of Poynting's 3-vector along a constant-time surface in each instantaneous reference frame is of no import to the transfer of energy-momentum, which must be evaluated using the fully relativistic tensor $T^{\mu \nu}$. To really make $\cur{R}$ zero, Poynting's vector would need to be null, in the instantaneous rest frame, on the light cone centered on the charge, and not on the space-like surface $t=0$.

Finally, we are left to prove the conservation of energy. If all of the external force $F_{\mathrm{ext}}^{\mu}$ in eq.\ \eqref{eq:lorentz} is transformed into kinetic energy, where does the radiated energy come from? We may answer this question by realizing that hyperbolic motion describes the hardly ``physical'' situation of a charge coming from and going to infinite $z$'s with an asymptotic speed approaching $c$. When we ask about the conservation of energy we are in fact trying to balance infinite quantities, and we should expect to do this, in some sense, ``in the limit''. Consider instead a trajectory built by attaching two portions of uniform motion to a uniformly accelerated motion taking place only through a finite lapse of time; at the junctions, the acceleration must necessarily be non-uniform. It is just there that radiation reaction acts to insure that the total work exerted by $F_{\mathrm{ext}}^{\mu}$ be equal to the increase in the kinetic energy of the charge \emph{plus} the energy radiated to infinity \cite{Ginzburg79,Tagliavini91}. We can picture the outward energy flux from the charge during uniformly accelerated motion as being borrowed from the divergent energy of the electro-magnetic field near the charge, which effectively acts as an infinite \emph{reservoir}. While draining energy, the field becomes more and more different from the velocity field of an inertially moving charge, and when hyperbolic motion finally ends, the external force will have to provide all the necessary energy to reestablish the original structure \cite{Tagliavini97}.

\subsection{Construction of a ``supported frame'' in a constant homogenous
gravitational field}
\label{subsec:frame}
Following Rohrlich \cite{Rohrlich63,Rohrlich65}, we shall seek a set of
non-inertial coordinates to ``simulate'' physics as seen by supported
experimenters. We will do so by appealing to their subjective observations
on basic facts about space-time and gravitation.

Supported observers follow a non-geodesic trajectory through Minkowski space-time; from their point of view, crono-geometry appears to be static (that is, the speed of clocks and the length of objects does not vary with time) and flat (space-time is Minkowskian); moreover, \emph{spatial} geometry appears as \emph{homogeneous} in the two ``horizontal'' directions. We shall then define a ``constant homogeneous gravitational field'' as a \emph{flat static metric that is manifestly invariant under translations and rotations in a spatial plane}. We will further impose the requirement that in these coordinates the geodesic equation reproduce the correct ``Newtonian behavior'' in the non-relativistic limit.

The most general metric satisfying such properties can be written as\footnote{See app.\ \ref{app:chfm}. Note that throughout the main text of this article we use primed letters to indicate coordinates and tensors in the supported frame; also, we denote the $D_t$ of the appendix as simply $D$.}
\begin{equation}
ds^2 = -D(z') \, {dt'}^2 + {dx'}^2 + {dy'}^2 +
( {\textstyle {\sqrt{D(z')} \,}' / g} )^2 \, {dz'}^2,
\label{eq:genmetric}
\end{equation}
where, because of the ``Newtonian limit'', $D(z')$ is required to approximate $1 + 2 g z'$ to first order in $g z'$. 
\emph{Supported observers} inhabit the worldlines of
constant $x'$, $y'$ and $z'$.
There are several explicit possibilities for $D(z')$,
stemming from the freedom to synchronize clocks differently at varying ``heights'' in the supported frame.
Consider for instance
\begin{equation}
ds^2 = -(1+2gz')\,{dt'}^2 + {dx'}^2 + {dy'}^2 + (1+2gz')^{-1} {dz'}^2,
\end{equation}
where $D(z')$ coincides with its non-relativistic limit. The most useful choice for $D(z')$, however, yields \emph{Rindler's metric}
\begin{equation}
\label{eq:rindler}
ds^2 = -(1+gz')^2{dt'}^2 + {dx'}^2 + {dy'}^2 + {dz'}^2,
\end{equation}
which implies a linear variation of clock speed by height: Einstein implicitly used this metric in his seminal argument about gravitational energy and the speed of clocks \cite{Einstein07}. Note that, independently of $D(z')$, it is always possible to put the metric in the Rindler form \eqref{eq:rindler} by introducing the new ``vertical'' coordinate $z''$ defined by $1 + g z'' = \sqrt{D(z')}$.

The transformation equations between the ``accelerated'' coordinates leading to \eqref{eq:genmetric} and the
inertial ones are given (up to Poincar\'e transformations) by
\begin{equation}
\left\{
\begin{aligned}
t &= g^{-1} \sqrt{D(z')} \sinh gt', \\
x &= x', \\
y &= y', \\
z &= g^{-1} \sqrt{D(z')} \cosh gt'.
\end{aligned}
\right.
\label{eq:accelcoord}
\end{equation}
From the inertial point of view, supported observers are seen to move on constantly accelerating hyperbolic trajectories (with accelerations depending on their $z'$). This is true also for supported charges, which will then radiate as discussed in \ref{subsec:uniform}, validating our prediction as to the result of experiment 3.

What about the converse? We obtain the trajectories of freely falling test bodies as seen from the supported frame by means of the geodesic equation in the ``accelerated'' coordinates (the required Christoffel coefficients may be found in eq.\ \eqref{eq:chrref} of app.\ \ref{app:chfm}).
The possibility of casting the metric in the unified Rindler form \eqref{eq:rindler} by a suitable choice of the coordinate system should not obscure the circumstance that for different choices of $D(z')$ we get different shapes for the geodesics, a fact that is obviously relevant to our considerations. In particular, hyperbolic motion is obtained only by setting \cite{Rohrlich63}
\begin{equation}
D(z') = \frac{1}{\cosh^2 \sqrt{(1-g z')^2 - 1}}.
\end{equation}
Interestingly, supported observers will not in general see a freely falling object move on a hyperbolic trajectory. As Rohrlich \cite{Rohrlich63} remarks, ``this provides a simple example dispelling the often expressed belief that in general relativity acceleration is relative and therefore reciprocal in the sense that the motion of A relative to B is identical (apart from a sense of direction) with the motion of B relative to A''.
\subsection{Physics in the supported frame}
\label{subsec:tests}
Let us put our supported reference frame to good use by calculating the result of experiments 1 and 2. Since the metric is not manifestly Minkowskian, we must explicitly incorporate it within Maxwell's equations of electro-magnetism:
\begin{gather}
{g'}^{\alpha \beta} \nabla_{\alpha} \nabla_{\beta} {A'}^{\mu} = -4 \pi
{j'}^{\mu},
\label{eq:maxwell} \\
{F'}_{\mu \nu} = \partial_{\mu} {A'}_{\nu} - \partial_{\nu} {A'}_{\mu},
\end{gather}
in the Lorentz gauge $\nabla_{\mu} {A'}^{\mu} = 0$. To model experiment 1, we must find the fields of a charge at rest at $x' = y' = z' = 0$ in the supported frame. Since however we know that the charge performs hyperbolic motion in the inertial frame, rather than solve equation \eqref{eq:maxwell} we may simply transform the field components \eqref{eq:fieldcomp}:
\begin{equation}
{F'}_{\mu \nu} =
\frac{\partial x^{\alpha}}{\partial {x'}^{\mu}}
\frac{\partial x^{\beta}}{\partial {x'}^{\nu}}
F_{\alpha \beta},
\end{equation}
and thus get \cite{Rohrlich63},
\begin{equation}
\begin{gathered}
E'_{\phi'} = H'_{\rho'} = H'_{\phi'} = H'_{z'} = 0, \\
E'_{\rho'} = g (z \, E_{\rho} - t \, H_{\phi}),
\hspace{0.5cm}
E'_{z'} = \frac{\mathrm{d} D(z')}{\mathrm{d}z'} \, \frac{E_z}{2 \, g};
\end{gathered}
\end{equation}
and then, using \eqref{eq:fieldcomp} and \eqref{eq:accelcoord},
\begin{equation}
\begin{gathered}
E'_{\rho'} = 8 \, e \, \rho' D(z') / g^3 {\xi'}^3, \\
E'_{z'} = - \frac{2 \, e \, ({\rho'}^2 + g^{-2} - g^{-2} D(z'))}{g^3 {\xi'}^3} \, \frac{\mathrm{d} D(z')}{\mathrm{d}z'}, \\
\xi' = g^{-2} \sqrt{(1 + g^2 {\rho'}^2)^2 + 2 (-1 + g^2 {\rho'}^2) D(z') + D(z')^2}.
\end{gathered}
\end{equation}
Because the magnetic field vanishes throughout space-time independently of $D(z')$, it is clear by our previous discussion that in the supported frame the charge is not radiating. Thus, even if there \emph{is} radiation in the inertial frame, any evidence of energy transfer is hidden to observers that are standing with the charge. What they see instead is an electric field which may be derived from the potential
\begin{equation}
\phi' = e g \frac{1 + g^2 {\rho'}^2 + D(z')}{[(1 - g^2 {\rho'}^2 - D(z'))^2 + 4 \, g^2 {\rho'}^2]^{1/2}}.
\label{eq:modifiedcoulomb}
\end{equation}
The shape of the field is dependent on the choice of $D(z')$: up to first order in $g$, and writing $D(z')$ up to second order as $D(z') = 1 + 2 z' g + \alpha {z'}^2 g^2 + O(g^3)$, it follows
\begin{equation}
\phi' = \frac{e}{r'} + \frac{e \, g \, z' ({\rho'}^2 - (\alpha - 2) {z'}^2)}{2 \, {r'}^3} + O(g^2),
\end{equation}
where obviously ${r'}^2 = {\rho'}^2 + {z'}^2$. 

The same reasoning may be applied to derive the fields of a freely falling charge by transforming a pure Coulomb field to the supported frame; the transformed fields turn out to be
\begin{equation}
\begin{gathered}
E'_{\phi'} = E'_{z'} = H'_{\phi'} = H'_{z'} = 0, \\
E'_{\rho'} = \frac{g \rho z \cdot E_{r}}{r},
\hspace{0.5cm}
E'_{z'} = \frac{\mathrm{d} D(z')}{\mathrm{d}z'}
\frac{z \cdot E_{r}}{2 \, g \, r},
\hspace{0.5cm}
H'_{\phi'} = - \frac{1}{D(z')} \frac{\mathrm{d} D(z')}{\mathrm{d}z'}
\frac{\rho t \cdot E_{r}}{2 \, r};
\end{gathered}
\label{eq:radfield}
\end{equation}
and using $E_{r}=e/r^2$ and \eqref{eq:accelcoord},
\begin{equation}
\begin{gathered}
E'_{\rho'} = \sqrt{D(z')} \frac{e \rho' \cosh g t'}{({\rho'}^2 + g^{-2} D(z') \cosh^2 g t')^{3/2}}, \\
E'_{z'} = \sqrt{D(z')} \, \frac{\mathrm{d} D(z')}{\mathrm{d}z'}
\frac{e \cosh g t'}{2 \, g^2 ({\rho'}^2 + g^{-2} D(z')^2 \cosh^2 g t')}, \\
H'_{\phi'} = - \frac{1}{\sqrt{D(z')}} \frac{\mathrm{d} D(z')}{\mathrm{d}z'} \frac{e \rho' \sinh g t'}{2 \, g ({\rho'}^2 + g^{-2} D(z') \cosh^2 g t')^{3/2}}.
\end{gathered}
\end{equation}
In analogy to the criterion outlined in section \ref{subsec:uniform} for the inertial case, we may evaluate the flux of the resulting energy-momentum tensor through light spheres centered on events on the worldline \cite{Rohrlich63}. This flux is non-null, proving that the charge does radiate.
This settles experiment 2 for good, and adds evidence to the claim that \emph{the notion of radiation is not invariant with respect to transformations from inertial to accelerated frames.} Yet we still have to clarify how this can be compatible with energy considerations.

\subsection{The balance of energy in the four experiments}
\label{subsec:energy}

While in both experiments 1 and 4 there is neither emission nor absorption of electro-magnetic energy to be accounted for, we are left to ascertain the source of the energy transferred to the detectors in experiments 2 and 3. The discussion of section \ref{subsec:uniform} applies directly to experiment 3: the energy for the radiation emitted by the supported charge and detected in the inertial frame must be provided by the agency enforcing the uniformly accelerated motion of the charge. This balance is apparent in the Dirac-Lorentz equation \eqref{eq:lorentz} for all ``physical'' accelerated motions, and stays true in the limit of pure hyperbolic motion.

Experiment 2 is the direct classical analogue of the Unruh effect: even if there is no detectable radiation in the inertial frame, the accelerated detector does absorb energy. In the accelerated frame this energy is seen to come from the radiation field \eqref{eq:radfield}. Yet, from the inertial point of view, the static Coulomb field of the freely falling charge has no energy to lose. If there were not an accelerating agency to support the detector, there would be no radiation to detect; hence, it is clear that all transferred energy must come from that agency.
One can argue as follows: any detector of electro-magnetic radiation must necessarily be charged on its own, and contain internal degrees of freedom. Thus the accelerating agency will have to supply an additional amount of work to balance the energy dissipated away by radiation reaction from the accelerating charged detector. Inertial observers will perceive this physical effect as a radiation field coming \emph{from the detector}.

Notice that this situation closely parallels what happens in the Unruh effect, where absorption of a ``Rindler particle'' by the accelerated detector is seen as emission of a ``Minkowski particle'' in the inertial frame \cite{Unruh84}. In the quantum case, this emission is due to the unavoidable coupling of the accelerated quantum detector to the vacuum state of the field. This coupling can be shown to justify classical radiation reaction via a fluctuation-dissipation theorem \cite{Sciama81,Callen51}.
\begin{appendix}
\section{Derivation of a constant homogeneous flat metric}
\label{app:chfm}
\noindent We reproduce here Rohrlich's derivation \cite{Rohrlich63} of the metric appropriate to a constant homogeneous gravitational field. We begin from the most general Lorentzian metric $g_{\mu \nu}$. Staticity implies that the time coordinate can be separated and the metric written as
\begin{equation}
ds^2 = -g_{tt} \, dt^2 + g_{ij} \, dx^i dx^j;
\end{equation}
where no coefficient depends on $t$. We can now diagonalize the spatial metric and impose homogeneity along coordinates $x$ and $y$: thus all coefficients will be functions of $z$ only,
\begin{equation}
ds^2 = -D_t(z) \, dt^2 + D_x(z) \, dx^2 + D_y(z) \, dy^2 + D_z(z) \, dz^2.
\end{equation}
The only non vanishing Christoffel coefficients turn out to be
\begin{equation}
\begin{gathered}
{\Gamma^{t}}_{tz} = \frac{1}{2} \frac{D_t'(z)}{D_t(z)}, \hspace{0.5cm}
{\Gamma^{x}}_{xz} = \frac{1}{2} \frac{D_x'(z)}{D_x(z)}, \\
{\Gamma^{z}}_{xx} = -\frac{1}{2} \frac{D_x'(z)}{D_z(z)}, \hspace{0.5cm}
{\Gamma^{z}}_{zz} = \frac{1}{2} \frac{D_z'(z)}{D_z(z)}, \hspace{0.5cm}
{\Gamma^{z}}_{tt} = \frac{1}{2} \frac{D_t'(z)}{D_z(z)}.
\end{gathered}
\end{equation}
We require flatness by imposing that all components of the Riemann tensor vanish. Thus we get the set of equations,
\begin{gather}
D_t' D_x' = D_t' D_y' = D_x' D_y' = 0; \\
2 D_i'' - \frac{(D_i')^2}{D_i} - \frac{D_z' D_i'}{D_z} = 0
\hspace{0.5cm} \text{for $i=x,y,t$;}
\label{eq:dts}
\end{gather}
which implies that two out of $D_t$, $D_x$ and $D_y$ must be constant. Let us now impose the Newtonian limit. The equation of motion for test particles falling in the gravitational field is ruled by the the geodesic equation:
\begin{equation}
\frac{d^2 x^{\rho}}{d \tau^2} + {\Gamma^{\mu}}_{\mu \nu} u^{\mu} u^{\nu} = 0.
\end{equation}
For motions much slower than the speed of light we may approximate the proper time $\tau$ with $t$ and the four-velocity $dx^{\mu}/d\tau$ as $(1,0,0,0)$. For the ``vertical'' component of the motion, we get
\begin{equation}
\frac{d^2 z}{d t^2} + {\Gamma^z}_{0 0} = \frac{d^2 z}{d t^2} + \frac{D_t'(z)}{2
D_z(z)}=0.
\label{eq:limit}
\end{equation}
From this equation, we learn that $D_t$ cannot be a constant, since otherwise we would not obtain a gravitational force field in the non-relativistic limit. Thus $D_x$ and $D_y$ must be constants that we can absorb in the definition of $x$ and $y$. Our new form for the metric is then
\begin{equation}
ds^2 = -D_t(z) \, dt^2 + dx^2 + dy^2 + D_z(z) \, dz^2,
\end{equation}
where by \eqref{eq:dts},
\begin{equation}
\frac{2 D_t''}{D_t'} - \frac{D_t'}{D_t} = \frac{D_z'}{D_z};
\end{equation}
hence,
\begin{equation}
D_z(z) = \left( C \frac{d}{dz} \sqrt{D_t(z)} \right)^2,
\end{equation}
where $C$ is a constant of integration. By \eqref{eq:limit}, for small
displacements $z$ we get $C=1/g$ and
\begin{equation}
D_t(z) \rightarrow 1 + 2 g z \hspace{0.5cm} \mathrm{for} \; gz \ll 1, gt \ll 1.
\end{equation}
We may therefore write our line element in the final form
\begin{equation}
ds^2 = - D_t(z) \, dt^2 + dx^2 + dy^2 + ( {\textstyle \sqrt{D_t(z)} \,}' / g )^2
dz^2,
\end{equation}
where $D_t(z)$ is required to approximate $1 + 2 g z$ to first order in $g z$. Finally we rewrite the updated non-null Christoffel coefficients,
\begin{equation}
{\Gamma^{t}}_{tz} = \frac{{\textstyle \sqrt{D_t(z)} \,}'}{\sqrt{D_t(z)}},
\hspace{0.5cm}
{\Gamma^{z}}_{zz} = \frac{{\textstyle \sqrt{D_t(z)} \,}''}{{\textstyle
\sqrt{D_t(z)} \,}'},
\hspace{0.5cm}
{\Gamma^{z}}_{tt} = g^2 \frac{\sqrt{D_t(z)}}{{\textstyle \sqrt{D_t(z)} \,}'}.
\label{eq:chrref}
\end{equation}
\end{appendix}
\providecommand{\bysame}{\leavevmode\hbox to3em{\hrulefill}\thinspace}

\end{document}